\documentclass[conference]{IEEEtran}

\usepackage{amssymb}
\usepackage{amsmath}
\usepackage{cite}
\usepackage{verbatim}
\usepackage{textcomp}
\usepackage[pdftex]{graphicx}
\usepackage{grffile}
\usepackage{caption}
\usepackage{subcaption}
\usepackage{epsfig}
\usepackage{epstopdf}
\usepackage{caption}
\usepackage[ruled,vlined,linesnumbered, commentsnumbered]{algorithm2e}
\usepackage{balance}
\usepackage{listings}
\usepackage{tikz}

\SetCommentSty{mycommfont}

\usepackage{tipa }	
\usepackage{pifont}	
\usepackage{wasysym}
\usepackage{color}

\usepackage{setspace}

\usepackage{amsthm}

\newtheorem{definition}{Definition}

\DeclareMathOperator*{\argmax}{arg\,max}
\DeclareMathOperator*{\argmin}{arg\,min}

\usepackage{color}

\begin{document}	

\title{Wireless Crowd Charging with Battery Aging Mitigation}

\author{
\IEEEauthorblockN{Tamoghna Ojha, Theofanis P. Raptis, Marco Conti, and Andrea Passarella}
\IEEEauthorblockA{Institute for Informatics and Telematics, National Research Council, Italy\\
Email: \small \{tamoghna.ojha, theofanis.raptis, marco.conti, andrea.passarella\}@iit.cnr.it
}}


\maketitle

\begin{tikzpicture}[remember picture,overlay]
\node[anchor=south,yshift=10pt] at (current page.south) {\fbox{\parbox{\dimexpr\textwidth-\fboxsep-\fboxrule\relax}{
  \footnotesize{
     \copyright 2022 IEEE. Personal use of this material is permitted.  Permission from IEEE must be obtained for all other uses, in any current or future media, including reprinting/republishing this material for advertising or promotional purposes, creating new collective works, for resale or redistribution to servers or lists, or reuse of any copyrighted component of this work in other works.
  }
}}};
\end{tikzpicture}

\begin{abstract}\label{Abst}
Battery aging is one of the major concerns for the pervasive devices such as smartphones, wearables and laptops. Current battery aging mitigation approaches only partially leverage the available options to prolong battery lifetime. In this regard, we claim that wireless crowd charging via network-wide smart charging protocols can provide a useful setting for applying battery aging mitigation. In this paper, for the first time in the state-of-the-art, we couple the two concepts and we design a fine-grained battery aging model in the context of wireless crowd charging, and two network-wide protocols to mitigate battery aging. Our approach directly challenges the related contemporary research paradigms by (i) taking into account important characteristic phenomena in the algorithmic modeling process related to fine-grained battery aging properties, (ii) deploying ubiquitous computing and network-wide protocols for battery aging mitigation, and (iii) fulfilling the user QoE expectations with respect to the enjoyment of a longer battery lifetime. Simulation-based results indicate that the proposed protocols are able to mitigate battery aging quickly in terms of nearly 46.74-60.87\% less reduction of battery capacity among the crowd, and partially outperform state-of-the-art protocols in terms of energy balance quality.
\end{abstract}



\section{Introduction}\label{Intro}

Battery technology development has progressed much more slowly compared to other areas of electronics, and, as a result, batteries are often considered the least green and the most design- and performance-limiting components of any electronic device \cite{armand2008,1251172}. Radical improvements in battery technology are rather unlikely, as their energy density is already very high. Incremental improvements are more likely, but some of those improvements will be eaten up by the energy demands of increased functionality in mobile devices. Since batteries are frequently the primary perishable energy source in portable devices, they can significantly affect the devices’ uptime and overall performance, and thus, the user’s experience \cite{7956309}. Recently, some manufacturers had to slow down smartphone development so as to account for aging batteries. It is therefore considered essential to explore alternative approaches (individualized or network-wide) to extend the battery performance \cite{1401835} and consequently the network lifetime.

The loss of battery capacity, known as battery aging \cite{Alma2020a}, is a big concern in pervasive user devices such as smartphones, wearables and laptops. Such pervasive devices which are becoming more functional with an increasing collection of resource-demanding applications, require higher energy loads and more expensive batteries. Therefore, a large part of the device price is incurred by the battery costs. Battery aging usually leads to premature replacements and disposal of entire, otherwise functional, devices. 

The battery lifetime is usually defined as the number of full charge–discharge cycles the battery can conduct before its end of life is reached, that is, its capacity reduces to less than a fraction of its initial capacity. There is a multitude of factors contributing to battery aging. A highly important one is the very high or very low state-of-charge (SOC). Many device users are not familiar with ways of improving the battery `health' of their devices, and therefore, frequently choose unnecessarily high SOC, which increases aging. As shown in \cite{7299844}, intelligently controlled actions, like usage-dependent decrease of charging levels, or charging at appropriate times can significantly mitigate battery aging.

Current battery aging mitigation approaches only partially leverage the available options to prolong battery lifetime. Specifically, for the first time in the state-of-the-art, we argue that pervasive device battery aging mitigation can be combined with wireless crowd charging \cite{8353364} via the possibilities that wireless power transfer \cite{7805205} can offer. By leveraging the functionality between device usage, and wireless crowd charging \cite{9183682}, the required charging levels for each device can be managed according to the battery lifetime objectives. By combining this functionality across multiple pervasive devices it can be possible to obtain a network-wide battery lifetime optimization, which can be used so as to configure near-optimal wireless charging profiles for each device. 

\subsection{Our contribution}

We argue that recent initial approaches which target individualized battery aging mitigation and user profiling, without taking into account neither other users in the networked population nor emerging energy sharing technologies like wireless crowd charging, are not uncovering the full potential of intelligent charging. Our approach aims at intelligently addressing (i) the battery aging technological problem, (ii) the
maximization of the user’s quality of experience (battery lifetime) socio-technical problem and (iii) characteristic network-wide energy distribution algorithmic problems, by not focusing on battery technology improvements per se, but on algorithmic, network-wide wireless charging optimization with battery aging mitigation. This vision directly challenges the related contemporary research paradigms, as follows:

\begin{enumerate}
\item Unlike mainstream, coarse-grained wired/wireless charge algorithmic modeling (such as \cite{Dhungana2019b}), which is characterized by simplifications in the representation of the underlying phenomena for the sake of decreasing the network modeling complexity (as highlighted in our previous work \cite{8271964}), our fine-grained \emph{modeling approach} takes into account important characteristic phenomena in the algorithmic modeling process related to fine-grained battery aging properties \cite{6974723}. Specifically, we, for the first time, model battery aging in the context of wireless crowd charging.
\item Unlike recent, mainstream battery aging mitigation visions (such as \cite{10.1145/3441644}) which focus on the optimization of individual devices/users for solely stationary wired charging, our approach designs and deploys \emph{ubiquitous computing} and network-wide protocols for battery aging mitigation for wireless crowd charging. We devise two such protocols which target to mitigate network-wide battery aging during wireless crowd charging.
\item Unlike mainstream crowd charging services which focus on solving emerging technical problems (such as \cite{8542781}), our approach inherently takes into account the fundamental human factor of the opportunistic setting and address the socio-technical dimension by fulfilling the user QoE expectations with respect to the \emph{enjoyment of a longer battery lifetime} via an aging mitigation enabled charging approach. Using simulations, we show that our approaches achieve performance trade-offs between battery aging mitigation (nearly 46.74-60.87\% less reduction of total battery capacity) and energy balance quality.
\end{enumerate}

The rest of the paper is organized as follows. In section \ref{sec:RelWrk}, we list the most representative literature on battery aging mitigation and wireless charging. Next, in Section \ref{sec:SysModel}, we present wireless energy transfer model and devise the detailed battery aging model in the context of wireless crowd charging. Sections \ref{sec:Bat_WEB} and \ref{sec:Result} discuss the proposed protocols and the performance evaluation results. Finally, the paper concludes in Section \ref{sec:Conclu} citing future directions of work.

\section{Related Works}\label{sec:RelWrk}

The current literature includes works solely in (i) the domain of battery aging mitigation for individual devices and without any wireless power capability, and, (ii) the domain of wireless crowd charging (usually targeting network energy balance) but without any battery aging mitigation mechanisms. The novelty of our approach lies on the combination of the two concepts for the first time. For completeness of the presentation, for each domain, we list here the most representative works in the literature so far.

\subsection{Battery aging mitigation for pervasive devices}

\cite{7298255} is probably the first paper that introduces, analyzes and mitigates the built-in battery aging when an individual device is operated with a provided power supply. The authors focus on the fact that in an effort to reduce size and weight, the capacity of the power supply is optimized for the average power demand rather than the maximum power demand.

In \cite{7299844}, the authors present Smart2, an advanced smartphone charger that mitigates battery's capacity fading, which until the introduction of the paper had usually been ignored. Smart2 exploits the fact that many users charge their phones over night. Since the overnight charging duration is unnecessarily long, the battery is subjected to a high average SOC, which accelerates battery aging. Therefore, Smart2 delays the charging process adaptively to be done shortly before the phone is unplugged.

In \cite{10.1145/3441644}, the authors present UBAR (User- and Battery-aware Resource management) which considers dynamic workload, user preference, and user plug-in/out pattern at run-time to provide a suitable trade-off between battery cycle life and quality of experience. UBAR personalizes this trade-off by learning the user’s habits, while considering battery temperature and (SOC) patterns. The evaluation results show that UBAR achieves 10\% to 40\% improvement compared to the existing state-of-the-art charging approaches.

\subsection{Wireless crowd charging}

In \cite{Nikoletseas2017}, the authors investigate how to efficiently transfer energy wirelessly in populations of battery-limited devices, towards prolonging their lifetime. They address a quite general case of diverse energy levels and priorities in the network and study the problem of how the network can efficiently reach a weighted energy balance state distributively, under both loss-less and lossy power transfer assumptions. Three protocols are designed, analyzed and evaluated, achieving different performance trade-offs between energy balance quality, convergence time and energy efficiency.

In \cite{Dhungana2019}, the authors try to minimize both the energy difference between nodes and the energy loss during this process. To this end, they propose three different energy sharing protocols between nodes based on different heuristics. One of the proposed protocols achieves the best performance by reaching an energy balance between nodes while keeping the maximum possible energy in the network.

In \cite{Ojha2021}, the authors present MobiWEB, a mobility-aware energy balancing method, which employs a predictor for estimating the mobility information of users. MobiWEB selects the different pairs of peers for energy exchange, such that the network energy distribution is balanced while minimizing the loss and energy difference between the peers. MobiWEB, when compared to the state-of-the-art, achieves different performance trade-offs between energy balance quality, convergence time, and energy-efficiency. 

In \cite{8422711}, the authors aim at finding the maximum number of wired charging times that could be skipped through utilization of available energy in other users in the vicinity with wireless energy sharing. To this end, they use dynamic programming approach to find the optimal skipping patterns for selfish and cooperative energy exchange cases and verify the results with brute force.

\section{Network Model}\label{sec:SysModel}

\subsection{Wireless Energy Transfer Model}
We consider $m$ number of users each carrying a smartphone with wireless charging facility in the area of interest. These users are denoted as $\mathcal{U} = \{ u_1, u_2, \cdots, u_m \}$, and the corresponding locations of these nodes at any time $t$ are represented as $\mathtt{L}(t) = \{l^1_t, l^2_t, \cdots , l^m_t\}$. The energy levels or SOCs of these nodes at any time $t$ are denoted by $\mathcal{E}_t = \{ E_t(1), E_t(2), \cdots, E_t(m) \}$. In this paper, we use the words `user' and `node', and `energy level' and `SOC', interchangeably. We assume that all the nodes are equipped with homogeneous wireless charging hardware. The energy transfer between any two nodes is affected by energy loss, and thus, the actual energy received by the receiving node is only a fraction of the transmitted energy. For example, the initial energy levels of nodes $u_i$ and $u_j$ nodes at time $t_1$ are $E_{t_1}(i)$ and $E_{t_1}(j)$, and $e$ energy is transferred from node $u_i$ to $u_j$ over a time duration ($t_1$,$t_2$). Then, the remaining energy of this pair of nodes at time $t_2$ will be, 
\begin{equation}
\Big( E_{t_2}(i), E_{t_2}(j) \Big) = \Big( E_{t_1}(i)-e, E_{t_1}(j)+(1-\beta)\times e \Big)
\label{eq:peer_rem_energy},
\end{equation}
where $\beta \in [0,1)$ denotes the energy loss factor during wireless charging. As typically considered in the related literature, such as \cite{Raptis2020} \cite{Bulut2020social}, we assume that $\beta$ remains constant over the duration of energy exchange. Also, any wireless charging exchange between two nodes does not affect the energy levels of other nodes ($\forall u_k \in \mathcal{U}, u_k \neq u_i, u_j$) in the network.

We define the energy distribution of all the nodes $\mathcal{U}$ at any time $t$, $\mathcal{E}_t(u)$, as,
\begin{equation}
\mathcal{E}_t(u) = \frac{E_t(u)}{E_t(\mathcal{U})} \qquad \mbox{ where, } E_t(\mathcal{U}) = \sum_{u \in \mathcal{U}} E_t(u)
\end{equation}
whereas, the average network energy is defined as, $\overline{E}_t = \frac{E_t(\mathcal{U})}{m}$.

We utilize the parameter energy variation distance to estimate the overall energy distribution in the whole network. The energy variation distance is calculated by applying the probability theory and stochastic processes described in \cite{ Nikoletseas2017, Dhungana2021}. For any two probability distributions $P$ and $Q$ defined over the sample space of $\mathcal{U}$, the total variation distance, $\mathcal{D}(P,Q)$, is computed as, 
\begin{equation}
\mathcal{D}(P,Q) = \sum_{x \in \mathcal{U}} |P(x) - Q(x)|
\label{eq:var_dist}
\end{equation}

As we consider the devices to be carried by the users, the movement pattern of these devices can be characterized by a human like mobility pattern. In our model, the users move from one location in the considered area to another following their own interest. Also, the movements of these users are such that they spend similar amount of time in the same location while revisiting it in a later point of time. Therefore, the movement of any user remains independent of other users as well. Subsequently, the number of users present in each location vary over time. 

We assume that the users can exchange energy wirelessly with any other user present in the same location at the same time. Specifically, such interaction between two users is only possible when the following conditions for a valid contact is satisfied. 
\begin{definition}
A valid contact ($\nu_{ij}^t$) between any two nodes $u_i$ and $u_j$ for a duration of $(t_1,t_2)$ satisfies the following conditions $\forall t \in (t_1,t_2)$,
\begin{equation} 
\nu_{ij}^t = 
\begin{cases}
1, \quad (\overline{l^i_t, l^j_t}) \leq d_{req} \mbox{ and } (t_2 - t_1) \geq t_{min},\\
0, \quad \mbox{otherwise}
\end{cases}
\end{equation}
where $(\overline{l^i_t, l^j_t})$ refers to the distance between $u_i$ and $u_j$. $d_{req}$ and $t_{min}$ denote the minimum distance and minimum time, respectively, required for performing a successful wireless energy exchange.
\end{definition}

\subsection{Battery Aging Model}
Any smartphone battery has a maximum number of charging/discharging cycles after which the battery's capacity drops to a specific fraction (for example, 80\%) of the initial capacity. We consider that after completing the maximum number of cycles, $\mathbb{C}_{max}$, the battery capacity is reduced by $\mathcal{P}_r$ percentage. For example, if the battery capacity after 500 cycles is reduced to 80\% (i.e., 20\% reduction), then $\mathbb{C}_{max} = 500$ and $\mathcal{P}_r = 20$. Also, for any user $u_i$, the parameter $\mathcal{C}_i(t)$ denotes the current number of completed battery cycles. According to battery specifications \cite{battery_guidelines}, one battery cycle refers to the period of use from fully-charged to fully-discharged and fully-charged again. To simplify the computations, in our work, we consider a battery cycle is completed when the cumulative charging and discharging percentages in recent charging and discharging sessions reach 100\%, respectively. Let, $\delta^c_{i,s}$ and $\delta^d_{i,s}$ denote the charging and discharging percentages, respectively, in each such charging/discharging session $s$ during the period $(t_1, t_2)$. Then, the number of current cycles of the battery is updated as,
\begin{equation}
\mathcal{C}_i(t_2) = \mathcal{C}_i(t_1) + 1   \quad \mbox{if } \sum_{s \in (t_1, t_2)} \delta^c_{i,s} = 100 \mbox{, } \sum_{s \in (t_1, t_2)} \delta^d_{i,s} = 100
\end{equation}

Next, we explain the different components of battery aging of any node participating in wireless crowd charging scenario. For any node $u_i$ at time $t$, the battery age depends on the number of already completed battery cycles ($\mathcal{C}_i(t)$). Subsequently, we can compute the degradation of the battery capacity till time $t$ as,
\begin{equation}
\mathcal{BC}_i^{curr} = \mathcal{C}_i(t) \times \frac{\mathcal{P}_r}{\mathbb{C}_{max}}
\label{eq:bh_1}
\end{equation}
where, $\frac{\mathcal{P}_r}{\mathbb{C}_{max}}$ denote the reduction of battery capacity for completing a battery cycle. 

Now, the other components for battery capacity is accounted based on whether the node is participating in the energy exchange or not. In case the node participates in energy exchange in session $s$ after time $t$, there will be a charging ($\delta^c_{i,s}$) or discharging ($\delta^d_{i,s}$) event depending on the status of the node as an energy receiver or provider, respectively. On the other hand, if the node does not participate in wireless charging, nonetheless, there will still be a discharging ($\delta^{no}_{i,s}$) depending on the usage of the device. In the following, we explain the computation of battery capacity corresponding to both these cases.

Using the definition of a battery cycle as given in \cite{battery_guidelines}, we can constitute a relationship between the charging/discharging percentage and battery cycle. Let, $x$ denotes the charging or discharging percentage in any session. Therefore, we can represent it as, $x = \delta^c_{i,s} + \delta^d_{i,s}$, as either of the two operations will be performed. For example, if the node $u_i$ is receiving energy from another node, then $\delta^c_{i,s} > 0$ and $\delta^d_{i,s} = 0$. Similarly, when $u_i$ is providing energy to another node, then $\delta^c_{i,s} = 0$ and $\delta^d_{i,s} > 0$. Subsequently, we can deduce the fraction of a battery cycle completed by the amount of charging/discharging in the recent session. Thus, for $x$, the equivalent fraction of battery cycle is $\frac{x}{2 \times 100}$, where the value $2 \times 100$ refers to the full battery cycle value, i.e., fully-discharged and fully-charged parts of a battery cycle. However, the charging and discharging process of any battery also varies with the battery's current capacity, i.e., any charging/discharging operation is more impactful as the overall capacity drops. We account this information in our computation by multiplying the battery capacity degradation with charging/discharging with the factor of elapsed battery capacity, i.e., $\frac{1 + \mathcal{C}_i(t)}{\mathbb{C}_{max}}$. Therefore, the change in battery capacity for participating in energy exchange is computed as,
\begin{equation}
\mathcal{BC}_i^{wcc} = \Big( \frac{\delta^c_{i,s} + \delta^d_{i,s}}{200} \times \frac{1 + \mathcal{C}_i(t)}{\mathbb{C}_{max}} \Big) \times \frac{\mathcal{P}_r}{\mathbb{C}_{max}}
\label{eq:bh_2}
\end{equation}

Similarly, when $u_i$ does not participates in energy exchange, the discharging of the battery by $\delta^{no}_{i,s}$ is equivalent to $\frac{\delta^{no}_{i,s}}{200}$ battery cycles. However, battery capacity also depends on the SOC and Depth-of-Discharge (DOD) -- very high and very low SOC (e.g. high DOD) both will lead to faster degradation of the battery capacity \cite{Alma2020a}. To reflect the same behavior in our model, we include a factor of $| 1 - \frac{E_t(i)}{50} |$ and multiply with the battery capacity degradation value. Here, $E_t(i)$ refers to the SOC value (in other word, energy level) of $u_i$ at time $t$, and $50$ refers to the middle point of the possible SOC values, i.e., 0--100. 
\begin{equation}
\mathcal{BC}_i^{nowcc} = \Big( \frac{\delta^{no}_{i,s}}{200} \times | 1 - \frac{E_t(i)}{50} | \Big) \times \frac{\mathcal{P}_r}{\mathbb{C}_{max}}
\label{eq:bh_3}
\end{equation}

Based on the information in Equation \eqref{eq:bh_1}, \eqref{eq:bh_2}, and \eqref{eq:bh_3}, we can compute the battery capacity after session $s$ at time $t$, $\mathcal{BC}_i(s,t)$, for any user $u_i$ as, 
\begin{equation}
\mathcal{BC}_i(s,t) = \mathcal{BC}_i^{curr} + \mathcal{I}_{wcc} \times \mathcal{BC}_i^{wcc} + \Big( 1 - \mathcal{I}_{wcc} \Big) \times \mathcal{BC}_i^{nowcc}
\end{equation}
where, $\mathcal{I}_{wcc}$ is the indicator variable which reflects whether the node is participating in energy exchange ($\mathcal{I}_{wcc} = 1$) or not ($\mathcal{I}_{wcc} = 0$).
 
Therefore, the battery capacity of $u_i$ is,
\begin{multline}
\mathcal{BC}_i(s,t) = \Big[ \mathcal{C}_i(t) + \mathcal{I}_{wcc} \times \Big( \frac{\delta^c_{i,s} + \delta^d_{i,s}}{200} \times \frac{1 + \mathcal{C}_i(t)}{\mathbb{C}_{max}} \Big) + \\
 \Big( 1 - \mathcal{I}_{wcc} \Big) \times \Big( \frac{\delta^{no}_{i,s}}{200} \times | 1 - \frac{E_t(i)}{50} | \Big) \Big] \times \frac{\mathcal{P}_r}{\mathbb{C}_{max}}
\end{multline}

\subsection{Problem Description}
In the following, we describe the studied problem in two parts -- considering energy balancing and considering battery aging. In the first case, towards the `energy balancing' goal, the focus is solely on achieving energy balance in the network without caring for the battery aging of the participating users. As discussed in \cite{Ojha2021} (our previous paper), this is a sensible goal in order to optimize network lifetime at the overall network level. We consider this problem as a benchmark when battery capacity degradation is not taken into account, and briefly summarize the main feature of the problem addressed in \cite{Ojha2021}. On the other hand, in the second case, we focus on mitigating the `battery aging' of the participating users without considering the energy balance goal for the network.

\subsubsection{Achieving Energy Balancing}\label{sec:sec:energy_bal_obj}
The energy balance of the network is achieved when the energy distribution of the nodes reach the same level. We name this stage as the target uniform energy distribution ($\mathcal{U}_T$). However, due to the inevitable energy loss during the energy exchange process, the actual energy distribution achieved at time $T$ will be different, i.e., $\mathcal{E}_T$. 

The final energy (SOC) level of any node $u_i$, $E_T(i)$, is calculated as,
\begin{equation}
E_T(i) = E_0(i) - \sum_{t \in T} \sum_{\forall u_j \neq u_i} e_{ij}^t + \sum_{t \in T} \sum_{\forall u_j \neq u_i} (1 - \beta)e_{ji}^t
\end{equation}
where, $E_0(i)$ refers to the initial energy (SOC) level of node $u_i$. $\sum_{t \in T} \sum_{\forall u_j \neq u_i} e_{ij}^t$ and $\sum_{t \in T} \sum_{\forall u_j \neq u_i} (1 - \beta)e_{ji}^t$ denote the energy transmitted to and received from other nodes $\forall u_j \neq u_i$, respectively.

Subsequently, the total energy loss is computed using the following,
\begin{equation}
\mathcal{L}_T = \sum_{t \in T} \sum_{\forall u_i \in \mathcal{U}} \sum_{\forall u_j \neq u_i} e_{ij}^t \beta
\end{equation}

In this part of the problem, our goal is to minimize the variation distance ($\mathcal{D}(\mathcal{E}_T, \mathcal{U}_T)$) between $\mathcal{E}_T$ and $\mathcal{U}_T$. Additionally, we need to minimize the total energy loss during the process of energy exchange.

\subsubsection{Mitigating Battery Aging}\label{sec:sec:bat_health_obj}
Considering battery aging of the users, irrespective of their participation in energy exchange, we need to minimize the total reduction in battery capacity for all the users in the network. However, considering the functionality of the network, we need to ensure that the nodes hold the minimum energy they need to continue the operations. Additionally, as discussed previously, higher energy of the nodes is not beneficial for the battery aging. Therefore, the SOC values of the nodes should ideally be within a range of desirable values. We count the number of nodes, which are not in that range, using the parameter, the number of unhealthy nodes, $\mathcal{N}_t^{UH}$. It is defined as,
\begin{equation}
\mathcal{N}_t^{UH} = |\mathcal{U}|_{E_{min} \geq E_t(i) \geq E_{max}}	\qquad \forall u_i \in \mathcal{U}
\end{equation}
where, $E_{min}$ and $E_{max}$ are the two threshold values which denote the minimum and maximum desirable SOC values for maintaining healthy battery. Subsequently, the total number of unhealthy nodes at time $T$ will be,
\begin{equation}
\mathcal{N}_T^{UH} = \sum_{t \in T} \mathcal{N}_t^{UH}
\end{equation}

Similarly, at time $T$, the total battery capacity of the network will be,
\begin{equation}
\mathcal{BC}_T = \sum_{t \in T} \sum_{s} \sum_{u_i \in \mathcal{U}} \mathcal{BC}_i(s,t)
\end{equation}

In this part of problem, our objective is to minimize the total number of unhealthy nodes as well as minimize the reduction in battery capacity ($\mathcal{BC}_T - \mathcal{BC}_0$) of the whole network over the duration $T$. It is important to note the following possible range of values for these parameters: $0 \leq \mathcal{C}_i(t) < \mathbb{C}_{max} $ and $0 < \delta^c_{i,s}, \delta^d_{i,s}, \delta^{no}_{i,s} < 100$.

\section{Battery Aging Mitigation and Wireless Crowd Charging}\label{sec:Bat_WEB}

\subsection{Wireless Energy Balancing}
In this part, we briefly explain the process followed to reach energy balancing in wireless crowd charging. In any ideal scenario, with no energy loss during energy exchange, at energy balance, all the nodes of the network will reach the average energy of the network. Let us consider this average energy value as $\overline{E}_t$. However, in practical scenarios, considering energy loss during energy exchange, the nodes will reach a different value, $\overline{E}^*$, rather than $\overline{E}_t$. $\overline{E}^*$ can be estimated using the value of energy loss ($\beta$) as shown in \cite{Dhungana2019},
\begin{equation}
\overline{E}^* = \frac{-(1-\beta) + \sqrt[2]{(1-\beta)}}{\beta} \qquad \forall \beta \in [0, 1], m \longrightarrow \infty
\end{equation}

Now, our objective is to achieve energy balance of the network with minimum energy loss and minimum energy variation distance of the population. Therefore, to achieve these two objectives, we need to select nodes with current energy level close to $\overline{E}^*$. In addition to this strategy, if we choose nodes with energy level at the opposite side of $\overline{E}^*$, we will be able to achieve higher decrease in the energy variation distance among the network in each iteration with lower energy loss. Therefore, using these two strategies, the nodes in the network will converge to energy balance state quickly. As the nodes move from one location to another, the inter-node meeting duration remains an important aspect to consider while choosing the peers of pairs to engage in wireless energy exchange. As mentioned, based on our two strategies, we start with the node (say, $u_i$) with energy level closest to $\overline{E}^*$. Therefore,
\begin{equation}
u_i = \argmin_{\forall u_i \in \mathcal{U}} |\overline{E}^* - E_t(i)|
\end{equation}

Then, we choose the designated pair for this node as the node (say, $u_j$) in the opposite side of $\overline{E}^*$, as well as energy level closest to $\overline{E}^*$. Thus, we apply the following conditions,
\begin{equation}
u_j = 
\begin{cases}
\argmin \limits_{\forall u_j \in N_t(u_i)} ( \overline{E}^* - E_t(j) ), \quad \mbox{ if, } E_t(i) > \overline{E}^* \\
\argmin \limits_{\forall u_j \in N_t(u_i)} ( E_t(j) - \overline{E}^* ), \quad \mbox{ otherwise} 
\end{cases}
\end{equation}
where, $N_t(u_i) = N_t(u_i) \cup u_j$ refers to the set of possible peers for $u_i$ at time $t$ and the condition $\nu_{ij}^t = 1$ holds.

Next, the selected pair of nodes engage in energy exchange, and we update their energy levels accordingly. We mark the nodes which have already reached the desired energy balance level. Thereafter, other pairs of nodes are selected from the nodes which are yet to achieve energy balance, and the process follows till the final time $T$. For further details of the protocol, intended readers can check \cite{Ojha2021}.

\begin{algorithm}[t!]
\caption{Battery Aging Mitigation and Peer Selection \textit{with} Prioritizing Network Availability ($P_{BA}$-$wNA$)}\label{algo:batweb_with_avl}
\textbf{Inputs:} $\mathcal{E}_t$, $\mathcal{BC}_0$.\\
\textbf{Output:} $\mathcal{N}_T^{UH}$, $\mathcal{BC}_T$.\\
Initialize $State[\cdot] \longleftarrow Incomplete$\;
\While{$t \leq T$}{
\For{$u_i \in \mathcal{U}$ and $State[i] = Incomplete$}{
	\If{$E_t(i) < E_{min}$}{
		Find the most unhealthy node, $u_i \longleftarrow \argmin_{\forall u_i \in \mathcal{U}} E_t(i)$\;
	}
	\For{$u_j \in \mathcal{U}$ and $u_j \neq u_i$}{	
		\If{$E_t(j) > E_{max}$}{
			Find the peer for $u_i$, $u_j \longleftarrow \argmax_{\forall u_j \in \mathcal{U}} E_t(j)$\;
		}
	}	
	Perform energy exchange between $u_i$ and $u_j$, $\Big( E_t(i), E_t(j) \Big) \longleftarrow \Big( E_t(i) + (1-\beta)\times \frac{E_t(j) - E_t(i)}{2}, E_t(j) - \frac{E_t(j) - E_t(i)}{2} \Big) $\;
	\If{$E_t(i) > E_{min}$}{
		Update $State[i] = Complete$\;
	}
	\If{$E_t(j) < E_{max}$}{
		Update $State[j] = Complete$\;
	}
	Update the number of unhealthy nodes $\mathcal{N}_t^{UH}$\;
	Update the battery capacity of nodes $\mathcal{BC}_i(\cdot)$ , $\mathcal{BC}_j(\cdot)$\;
}
}
\end{algorithm}

\subsection{Battery Aging Mitigation}
In this part, we discuss our proposed battery aging mitigation protocol for wireless crowd charging. As mentioned in Section \ref{sec:sec:bat_health_obj}, our objective is to minimize the number of unhealthy nodes as well as minimize the reduction of overall battery health of the whole network. To achieve these objectives, we devise a greedy protocol which selects the `unhealthy' nodes and pair them with suitable `unhealthy' nodes for energy exchange. In this way, these nodes return to the `healthy' region again. Now, we can select the unhealthy nodes in two different ways -- based on the priority to ensure network availability or not. To ensure network availability, our focus will be to prioritize recharging the nodes which are about to get their batteries exhausted.

\subsubsection{Prioritizing Network Availability}
In the case of prioritizing the network availability, we need to select the nodes having current SOC closest to zero first. So, we select the node $u_i$ as,
\begin{equation}
u_i = \argmin_{\forall u_i \in \mathcal{U}} E_t(i)	\qquad E_t(i) < E_{min}
\end{equation}

Thereafter, we need to select $u_j$, the peer of $u_i$ for energy exchange, from the other `unhealthy' zone where $E_t(i) > E_{max}$. Therefore, 
\begin{equation}
u_j = \argmax_{\forall u_j \in \mathcal{U}} E_t(j)	\qquad E_t(j) > E_{max}
\end{equation}

\begin{algorithm}[t!]
\caption{Battery Aging Mitigation and Peer Selection \textit{without} Prioritizing Network Availability ($P_{BA}$-$woNA$)}\label{algo:batweb_without_avl}
\textbf{Inputs:} $\mathcal{E}_t$, $\mathcal{BC}_0$.\\
\textbf{Output:} $\mathcal{N}_T^{UH}$, $\mathcal{BC}_T$.\\
Initialize $State[\cdot] \longleftarrow Incomplete$\;
\While{$t \leq T$}{
\For{$u_i \in \mathcal{U}$ and $State[i] = Incomplete$}{
	\If{$E_t(i) < E_{min}$}{
		Find the unhealthy node, $u_i \longleftarrow \argmin_{\forall u_i \in \mathcal{U}} \mbox{  } min( E_{min} - E_t(i), E_t(i) - E_{max} )$\;
	}
	\For{$u_j \in \mathcal{U}$ and $u_j \neq u_i$}{	
		\If{$E_t(i) < E_{min}$ and $E_t(j) > E_{max}$}{
			Find the peer for $u_i$, $u_j \longleftarrow \argmax_{\forall u_j \in \mathcal{U}} E_t(j) - E_{max}$\;
		}
		\If{$E_t(i) > E_{max}$ and $E_t(j) < E_{min}$}{
			Find the peer for $u_i$, $u_j \longleftarrow \argmax_{\forall u_j \in \mathcal{U}} E_{min} - E_t(j)$\;
		}
	}	
	Perform energy exchange between $u_i$ and $u_j$\; 
	\If{$E_t(i) < E_{min}$ and $E_t(j) > E_{max}$}{
	    $\Big( E_t(i), E_t(j) \Big) \longleftarrow \Big( E_t(i) + (1-\beta)\times \frac{E_t(j) - E_t(i)}{2}, E_t(j) - \frac{E_t(j) - E_t(i)}{2} \Big) $\;
	}    
	\If{$E_t(i) > E_{max}$ and $E_t(j) < E_{min}$}{
	    $\Big( E_t(i), E_t(j) \Big) \longleftarrow \Big( E_t(i) - \frac{E_t(j) - E_t(i)}{2}, E_t(j) + (1-\beta)\times \frac{E_t(j) - E_t(i)}{2} \Big) $\;
	}
	\If{$ E_{min} < E_t(i) < E_{max} $}{
		Update $State[i] = Complete$\;
	}
	\If{$ E_{min} < E_t(j) < E_{max} $}{
		Update $State[j] = Complete$\;
	}
	Update the number of unhealthy nodes $\mathcal{N}_t^{UH}$\;
	Update the battery capacity of nodes $\mathcal{BC}_i(\cdot)$ , $\mathcal{BC}_j(\cdot)$\;
}
}
\end{algorithm}

\subsubsection{Without Prioritizing Network Availability}
In this scenario, we start with the unhealthy node which is having current SOC level closest to the energy threshold levels. Therefore, we select the first node, $u_i$ as,
\begin{equation}
u_i = \argmin_{\forall u_i \in \mathcal{U}} \mbox{  } min( E_{min} - E_t(i), E_t(i) - E_{max} )
\end{equation}
where, $min(a, b)$ return the minimum value among $a$ and $b$.

Thereafter, we choose $u_j$ from the nodes which has current SOC values in the other `unhealthy' zone compared to $u_i$. We apply similar strategy to choose the node which has the energy level closest to the respective SOC threshold value. Therefore,
\begin{equation}
u_j = 
\begin{cases}
\argmin \limits_{\forall u_j \in \mathcal{U}; E_t(j) > E_{max}} E_t(j) - E_{max}, \mbox{ if, } E_t(i) < E_{min} \\
\argmin \limits_{\forall u_j \in \mathcal{U}; E_t(j) < E_{min}} E_{min} - E_t(j), \mbox{ if, } E_t(i) > E_{max}
\end{cases}
\end{equation}

Subsequently, we apply these conditions on the rest of nodes of the network until there are no such unhealthy nodes or not enough available peers. Algorithm \ref{algo:batweb_with_avl} and \ref{algo:batweb_without_avl} list the steps followed by the nodes of the network with and without the priority for network availability, respectively.

\textit{Complexity of the Algorithms}: Both the Algorithms \ref{algo:batweb_with_avl} and \ref{algo:batweb_without_avl} have an execution time of $O(m^2)$, where $m$ denotes the number of users.

\section{Performance Evaluation}\label{sec:Result}

\subsection{Simulation Settings}
We consider 100 nodes deployed randomly over 5 different locations. In each iteration, the nodes' movement is as follows -- they choose a random stay duration between 10-30 $minutes$, and after that time, the nodes move to another location chosen randomly. The initial energy distribution or the SOC values of the nodes are chosen randomly over $[0, 100]$ units. We also consider that the current completed battery cycles ($\mathcal{C}_i(t)$) of these nodes are randomly distributed over $[0, 0.5 \times \mathbb{C}_{max}]$. The nodes are equipped with homogeneous wireless charging hardware with energy loss rate $\beta = 0.2$. We assume the wireless charging rate $\alpha = 0.5$ to simulate a real Qi charger (for example, the Qi charger in \cite{Belkin_qi} with capacity of 7.5 $Wh$ has $\alpha \approx 0.675$). In the experiments, we consider $E_{min}$ and $E_{max}$ as 20 and 80, respectively. We assume $\mathcal{P}_r$ and $\mathbb{C}_{max}$ as 20 and 500, respectively, based on a Lithium Polymer battery Pack of 4000 $mAh$ capacity \cite{Battery_4000}.

\begin{figure*}[ht]
        \centering
        \begin{subfigure}[b]{0.25\textwidth}
                \includegraphics[width=\textwidth]{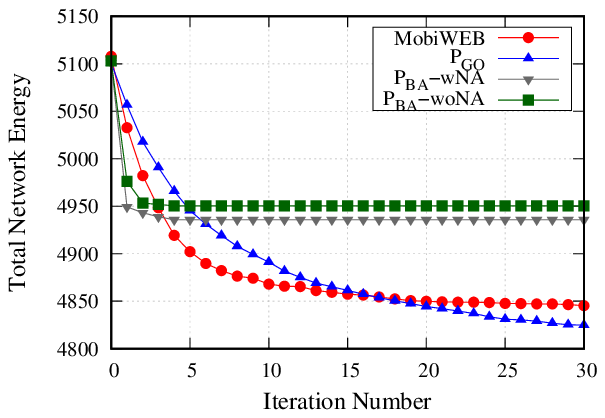}
                \caption{Total network energy.}
                \label{fig:tot_energy}
        \end{subfigure}%
        \begin{subfigure}[b]{0.25\textwidth}
                \includegraphics[width=\textwidth]{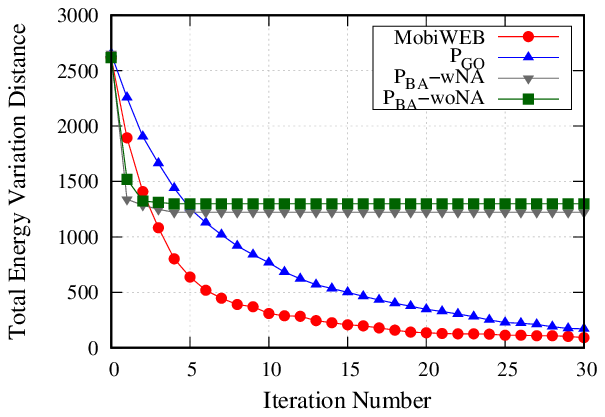}
                \caption{Total energy variation distance.}
                \label{fig:tot_var_dist}
        \end{subfigure}%
        \begin{subfigure}[b]{0.25\textwidth}
                \includegraphics[width=\textwidth]{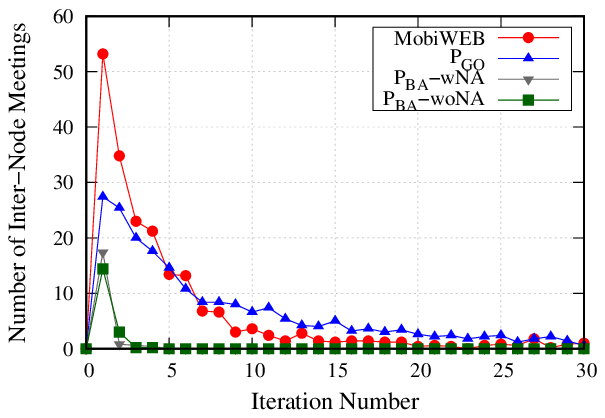}
                \caption{Number of inter-node meetings.}
                \label{fig:num_meet}
        \end{subfigure}%
        \begin{subfigure}[b]{0.25\textwidth}
                \includegraphics[width=\textwidth]{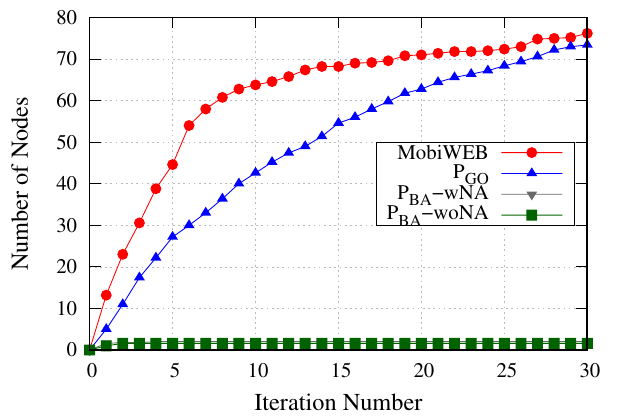}
                \caption{Nodes reached energy balance.}
                \label{fig:num_complete}
        \end{subfigure}%
        \caption{Performance comparison w.r.t. energy balance quality.}
\end{figure*}

\subsection{Benchmarks and Evaluation Metrics}
We consider two recent state-of-the-art methods $MobiWEB$ \cite{Ojha2021} and $P_{GO}$ \cite{Dhungana2019} as the benchmarks for comparing the performance of our proposed methods. In both the benchmarks, first, a pair of nodes are selected with current SOC level that is closest to the target energy balance level as well as belong in the opposite side of it. Subsequently, more pairs of peers are selected from the rest of the nodes for energy exchange. However, $MobiWEB$ leverages the user mobility information, and accordingly fine-grains the selection of nodes for network-wide wireless crowd charging. Whereas, $P_{GO}$ chooses nodes solely based on energy conditions rather that joint conditions of mobility and SOC. Our proposed protocols, on the other hand, select nodes based on the SOC levels and the priority of network availability. In all the protocols, we consider energy loss and the energy exchange duration is bounded by the inter-node meeting duration of the participating nodes. However, as only $MobiWEB$ leverages mobility prediction, the comparison is a bit unfair to the proposed protocols.

To compare the performance of the proposed and benchmark methods, we show the results using the metrics which cover both energy balance as well as the battery aging mitigation aspects. The metrics related to energy balance are total network energy, total energy variation distance, number of inter-node meetings, and number of nodes that reached energy balance. With respect to battery aging mitigation, we consider the metric: reduction of total battery capacity, which refers to the difference of total battery capacity of the crowd between the iterations ($\mathcal{BC}_T - \mathcal{BC}_0$).  

\subsection{Results}

\subsubsection{Total network energy}
Figure \ref{fig:tot_energy} shows the results for total network energy (SOC levels) in different iterations. Among $MobiWEB$ and $P_{GO}$, $MobiWEB$ experiences higher energy loss due to increased number of inter-node meetings during the initial parts of the experiment (i.e. 1-5). This behavior is supported by the results shown in Figure \ref{fig:num_meet}. Subsequently, the total network SOC reduces quickly compared to $P_{GO}$. However, the difference in total network energy reduces significantly in the higher iterations (i.e. 20-30) compared to the initial iterations (1-10). On the other hand, for $P_{BA}$-$wNA$ and $P_{BA}$-$woNA$, only specific nodes meeting the conditions of SOC threshold levels are chosen for the energy exchange, and most of these nodes are chosen in the initial iterations (1-3). Accordingly, the energy losses happen during these iterations, and the total energy level does not change afterwards. However, the energy losses are higher in both these protocols due to the higher amount of energy exchanged (compared to $MobiWEB$ and $P_{GO}$) between the selected nodes. Among $P_{BA}$-$wNA$ and $P_{BA}$-$woNA$, the energy loss in $P_{BA}$-$wNA$ is higher compared to $P_{BA}$-$woNA$. Such behavior is attributed to the type of peer selection promoted in $P_{BA}$-$wNA$ -- energy exchange between peers with very high or very low SOC results in higher amount of energy exchanged and energy loss.

\subsubsection{Total energy variation distance}
The results for total energy variation distance is depicted in Figure \ref{fig:tot_var_dist}. From the results it is evident that the higher amount of energy exchange during the initial iterations (1-3) in $P_{BA}$-$wNA$ and $P_{BA}$-$woNA$ helps in reducing the total energy variation distance among the crowd very early. In subsequent iteration, as there are not much energy exchange activity in these two protocols, the variation distance remains unchanged. Similarly, higher number of inter-node interactions in $MobiWEB$ results in lower energy variation distance compared to $P_{GO}$. Although the proposed protocols exhibit lower total energy variation distance, the energy loss remains higher compared to the benchmarks. However, we also notice that both the proposed protocols are able to exploit the possible inter-node meeting opportunities in the early iterations, as shown in Figure \ref{fig:num_meet}.

\subsubsection{Number of inter-node meetings}
We depict the number of inter-node interactions during the experiments in Figure \ref{fig:num_meet}. It is depicted from the results that the proposed protocols are able to explore higher percentage of total possible meeting opportunities ($> 90\%$) compared to the benchmarks ($35-55\%$) during initial parts of experiments (iterations 1-3). However, the absolute number of inter-node meetings remains higher in the benchmarks. Such behavior is attributed to the higher number of potential peers in the benchmarks compared to few number of potential peers in case of proposed protocols.

\subsubsection{Number of nodes that reached energy balance}
In Figure \ref{fig:num_complete}, we plot the number of nodes with SOC equal to the target energy balance level. The results show that very less number of nodes reach the desired SOC level in the proposed protocols compared to the benchmarks. The reason for such performance is due to the type of energy exchange promoted in the proposed protocols -- only nodes with SOC $E_t(\cdot) < E_{min}$ and $E_t(\cdot) > E_{max}$ exchange energy without regarding the energy balance level. Therefore, although most of the inter-node meeting opportunities are exploited and the proposed protocols are able to reduce the energy variation distance, the number of nodes that reach the energy balance level remains lower.

\begin{figure}[ht]
\centering
\includegraphics[scale=0.65]{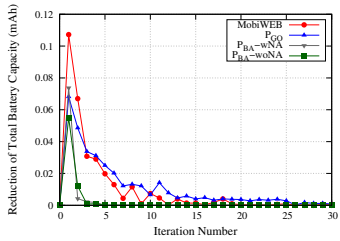}
\caption{Reduction in total battery capacity.}
\label{fig:tot_capacity}
\end{figure}


\subsubsection{Reduction of total battery capacity}
We plot the reduction of the total battery capacity at different iterations in Figure \ref{fig:tot_capacity}. From the results it is evident that the reduction of the total battery capacity for the whole crowd is significantly lower in the proposed protocols compared to the benchmarks. To be specific, the cumulative (over iterations 1-3, where the comparison is fair) reduction in total battery capacity is nearly 46.74-60.87\% less in proposed protocols compared to the benchmarks. Therefore, we can infer that on a long run, the proposed protocols will be able to reduce battery aging (or extend the network lifetime) significantly compared to the benchmarks. In $P_{BA}$-$wNA$ and $P_{BA}$-$woNA$, the reduction of total battery capacity remains lower, and such performance is attributed to the lower number of inter-node interactions which results in lower energy loss as well. However, the energy loss in $P_{BA}$-$wNA$ is higher compared to $P_{BA}$-$woNA$ due to the higher amount of energy exchanged between the selected peers  in $P_{BA}$-$wNA$. This behavior is attributed to the type of peer selection promoted in $P_{BA}$-$wNA$: energy exchange between peers with very high or very low SOC results in higher amount of energy exchanged and energy loss.


\section{Conclusion}\label{sec:Conclu}
In this paper, we, for the first time, model battery aging in the context of wireless crowd charging. Thereafter, we present two protocols, namely $P_{BA}$-$wNA$ and $P_{BA}$-$woNA$, to mitigate network-wide battery aging during wireless crowd charging. The proposed protocols enable the selection of pairs for energy exchange such that the battery lifetime is enhanced. Simulation results depict that the proposed protocols are able to mitigate battery aging quickly in terms of nearly 46.74-60.87\% less reduction of battery capacity among the crowd. Regarding energy balance quality, proposed protocols are able to outperform the benchmarks partially during the experiments. Therefore, we can conclude that the proposed protocols achieve performance trade-offs between battery aging mitigation and energy balance quality. The future works will be focused on improving the protocols further on the joint objectives of battery aging mitigation and energy balance quality.

\section*{Acknowledgment}
This work was carried out during the tenure of an ERCIM `Alain Bensoussan' Fellowship Programme of the first author. This work was partially funded by the European Union's Horizon 2020 Research and Innovation Programme under grant agreement No. 951972 (StandICT.eu 2023) through its 4th open call.

\balance
\bibliographystyle{IEEEtran}
\bibliography{BatWEB}

\end{document}